\documentclass{webofc}
\usepackage[varg]{txfonts}   % Web of Conferences font
\begin{document}
\title{Moving the California distributed CMS xcache from bare metal into containers using Kubernetes}

\author{\firstname{Edgar} \lastname{Fajardo}\inst{1}\fnsep\thanks{\email{emfajard@ucsd.edu}} \and
        \firstname{Matevz} \lastname{Tadel}\inst{1}\fnsep\thanks{\email{mtadel@ucsd.edu}} \and
        \firstname{Justas} \lastname{Balcas}\inst{2}\fnsep\thanks{\email{jbalcas@caltech.edu}} \and
        \firstname{Alja} \lastname{Tadel}\inst{1}\fnsep\thanks{\email{amraktadel@ucsd.edu}} \and
        \firstname{Frank} \lastname{Wuerthwein}\inst{1}\fnsep\thanks{\email{fkw@ucsd.edu}} \and
        \firstname{Diego} \lastname{Davila}\inst{1}\fnsep\thanks{\email{didavila@ucsd.edu}} \and
        \firstname{Jonathan} \lastname{Guiang}\inst{1}\fnsep\thanks{\email{jguiang@ucsd.edu}} \and
        \firstname{Igor} \lastname{Sfiligoi}\inst{1}\fnsep\thanks{\email{isfiligoi@sdsc.edu}}
        % etc.
}

\institute{9500 Gilman Dr, La Jolla, CA 92093
\and
           1200 E California Blvd, Pasadena, CA 91125 
          }

\abstract{%
  The University of California system has excellent networking between all of its campuses as well as a number of other Universities in CA, including Caltech, most of them being connected at 100 Gbps. UCSD and Caltech have thus joined their disk systems into a single logical xcache system, with worker nodes from both sites accessing data from disks at either site. This setup has been in place for a couple years now and has shown to work very well. Coherently managing nodes at multiple physical locations has however not been trivial, and we have been looking for ways to improve operations. With the Pacific Research Platform (PRP) now providing a Kubernetes resource pool spanning resources in the science DMZs of all the UC campuses, we have recently migrated the xcache services from being hosted bare-metal into containers. This paper presents our experience in both migrating to and operating in the new environment.
}
\maketitle
\section{Introduction}
\label{intro}
As part of the preparations for the High Luminosity Large Hadron Collider (HL-LHC) new ventures on data access and data placing have been taken place. In the past the Anydata, anytime, anywhere (AAA) \cite{cmsDeployAAA}\cite{xrootdcms} project showed that the CMS Computing model could be changed towards a model in which the location of the data could be separated from the location of the computing resources. Since data could be accessed from the Wide Area Network (WAN) then where to place data and how to place it for remote and local access became an important question. However if the same data is read many times from close enough resources efficiencies can be gained from caching the subsequent WAN reads, hence an XRootD \cite{dorigo2005xrootd} based caching technology was born: XCache \cite{xcache}. XRootD servers allow for a tree based structure called a federation in which the bottom level leaves are servers holding data and the upper level leaves are called redirectors. Once a client asks for a file to a redirector this one asks its children if they have such file if one has it the client is redirected to that server if none have it the client is redirected to a leaf above for the same process to happen again. However this implies quite some latency behind file opens to servers geographically very far. The XCache hides these latencies as it sits between a client and an XRootD federation and based on demand it fetches files from the federation and stores them locally. For the CMS Run II, UCSD scale tested and deployed into production a multi node Federated XCache that could serve a limited part of the analysis namespace \cite{FederatedCache}.

In this project we show how two grid sites (Caltech and UCSD) deployed in production a combined multi site logical XCache, how data analytic drove the hardware specifications and specification of namespace, and finally how Kubernetes was chosen to allow for fast deployments of this newly built infrastructure.

\section{Namespace and Working Set}
\label{namespace}
\begin{figure}[h]
\centering
\sidecaption
\includegraphics[width=8cm,clip]{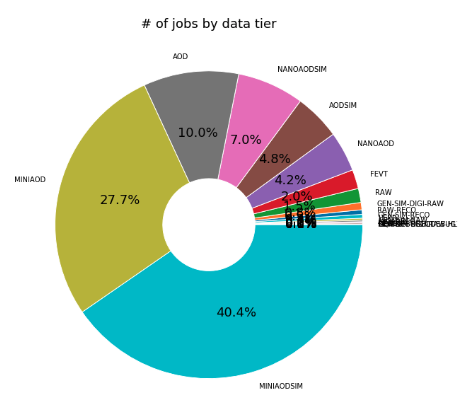}
\caption{Distribution of Analysis Jobs in CMS in 2019}
\label{jobs-miniaod}       % Give a unique label
\end{figure}

The driving cost for deploying a cache is the disk. So the most efficient use of resources in a cache is to choose the namespace in away that the maximum amount of jobs can use it while using the least amount of space. Using the CERN based analytics database of past CMS data access we concluded that the namespace of choice would be MINIAOD. From figure \ref{jobs-miniaod} we concluded that seventy percent of CMS jobs used the MINIAOD format.
%TODO: ADD a reference to the CERN hadoop

If we were to cache all the MINIAOD and MINIAODSIM data from CMS in the cache, according to table \ref{tab-datasetsize} we would need approximately 7.5 Petabytes of disk. However we envisioned that we would not need to provision space for all the datasets but only for the working set hence a more efficient use of the disk. 

    \begin{table}
\centering
\caption{Size of all datasets in different MINIAOD Data tiers}
\label{tab-datasetsize}       % Give a unique label
% For LaTeX tables you can use
\begin{tabular}{c | c }
\hline
\textbf{Data Tier} & Size (PB)\\\hline
 /*/*/MINIAOD & 2.92   \\
/*/*/MINIAODSIM & 4.6 \\\hline
\end{tabular}
\end{table}

We defined the working set as the size of the unique set of files from the the MINIAOD and MINIAODSIM datatiers accessed by the analysis jobs through the SoCal cache. This is calculated by creating a list of all the files accessed within a month then dropping the duplicates i.e., a file that is accessed N time within a month will only appear once in the list, afterwards we get the file size for every file on the list and finally we add up all these sizes. The working set will tell us how much disk is needed to store all the accessed files within a given month. In order to have both an upper bound and a close approximation to our use case we calculated this metric using two different approaches:
\begin{itemize}
    \item \textbf{Global working set}: takes into account the accesses in all of the sites. It is calculated using the data of Global Pool \cite{globalpool} monitoring system
    \item \textbf{SoCal working set}: only considers the accesses at the UCSD and Caltech sites. It is calculated using the Xrootd monitoring data of the SoCal Cache \cite{FederatedCache} deployment
\end{itemize}

These two different monitoring data sources present different data access granularities (dataset vs file). In the case of the global pool monitoring a record stores the dataset of the file being accessed whereas in the Xrootd, the actual file being accessed is what is stored. This granularity plays an important role when the working set is calculated. When using the dataset granularity an access to a file in a given dataset will add up the entire size of the dataset to the working set whereas in the case of the file granularity it would only add up that specific file size. Even when is common that an analysis task will access an entire dataset this is not always true, e.g., in the case of jobs reading Monte Carlo data they will only read parts of the dataset until they have enough (statistical) samples.

\begin{figure}[h]
\centering
\sidecaption
\includegraphics[width=8cm,clip]{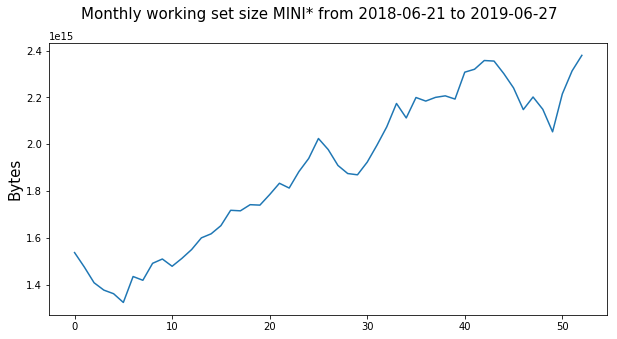}
\caption{Monthly global working set size for analysis jobs using MINIAOD and MINIAODSIM data tiers}
\label{workingset-method1}       % Give a unique label
\end{figure}

\begin{figure}[h]
\centering
\sidecaption
\includegraphics[width=8cm,clip]{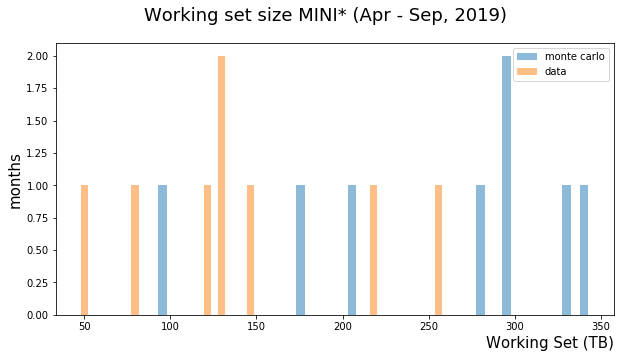}
\caption{Monthly SoCal working set size for analysis jobs using MINIAOD and MINIAODSIM data tiers}
\label{workingset-method2}       % Give a unique label
\end{figure}

In figure \ref{workingset-method1} we can see that the global working set goes from 1.2PB to 2PB over the course of a year, each data point shows the size of the working set of the 4 previous weeks. In figure \ref{workingset-method2} we see the SoCal working set, a bar in this histogram shows the number of months in which the working set size, at the x-axis, has been observed. In this case the accesses to detector data and Monte Carlo generated data are separated so they need to be added up to show the total working set size, as an example, for October 2019 we got a total working set size of 451TB.

\section{Hardware Provisioning and Deployment}
Based on the calculations of section \ref{namespace} we determined the amount of disk space that needed to be provisioned for these efforts. The current hardware deployment is summarized in Table \ref{tab-hardware}. However given the fact that the working set seems to be consistently growing (see figure: \ref{workingset-method1}) it was clear we need a way to easily deploy new caches in heterogeneous hardware. 

\begin{table}
\centering
\caption{Hardware deployment specification for SoCal cache}
\label{tab-hardware}       % Give a unique label
% For LaTeX tables you can use
\begin{tabular}{c | c |c}
\hline
 & \textbf{UCSD} & \textbf{Caltech}   \\\hline
\textbf{Nodes} & 11 (+1 JBOD\footnotemark) & 2 \\
\textbf{Disk Capacity per node} & 12 x 2TB = 24TB (+ 48 x 11TB) & 30 x 6TB = 180TB
 \\ 
\textbf{Network Card per node} & 10 Gbps (+ 40 Gbps) & 40 Gbps\\
\textbf{Total Disk Capacity} & 264 TB (+ 528 TB) = 792 TB & 360 TB* \\\hline
\textbf{Total} & \multicolumn{2}{|c}{$792 TB + 360 TB = 1,152 TB$} \\

\end{tabular}
\end{table}
\footnotetext{JBOD stands for Just a bunch of disks. Xrootd stores data on each of them separately without any raid setup. This is acceptable for caching since if one disk goes bad the cache is still available and the data loss can be retrieved after from the federation}

\subsection{Kubernetes}
Given the heterogeneous type of systems in which caches are deployed, the fact they are located in several institutions and the rapid development of new features of XrootD we took decision to host to operate newer caches on a DevOps model. The ideas is to have quick development and deployment in production without the intervention of the system administrators of the bare metal.  We pick Kubernetes as the tool to ease deployment of the XCache services. The Open Science Grid was already building the newest version of RPM coming from XrootD developers hence it was friction-less for them to build Docker containers out of each new version. Moreover we build specialized containers for the CMS XCache deployments on top of the XrootD OSG generic caches.

Right now we are able to deploy a new containers in less than a minute in production and rolled them back just as fast. We also took advantage of several Kubernetes functionalities like secrets and volume claims. The first one (secrets) to manage the certificates of each cache (The one it needs to authenticate to the Xrootd Federation and the ones to authenticate with the clients) and the latter one to allow for other pods to run in same bare metal but not been able to use the the disks (or partitions) we claimed. This potentially allows (although we have not tried it yet) to let users of the kubernetes federation use the extra resources the cache pod is not using on the bare metal host. 

There are indeed some potential frictions with this deployment model: at its core the container service deployment is based on a stateless deployment of micro services and the cache by its nature is stateful (the state is the cached files on local disk), the volume claims (local)allow us to bridge this gap. Since the state is only in local cached files it allows us to rollback and forward container versions with total ease and in the worst case scenario a cache (given its nature of having a non custodial copy) can always be repopulated on demand.

To obtain the best performance we did not use the container virtualized network. Instead we opted for our cache pods to bind directly to the hosts network. For future work we would like to understand what is the performance hit of using this capabilities. Right now we do not see any performance difference between containerized deployments versus bare metals, but this is a direction of future work.

\begin{figure}[h]
% to insert the figure file..
\centering
\sidecaption
\includegraphics[width=8cm,clip]{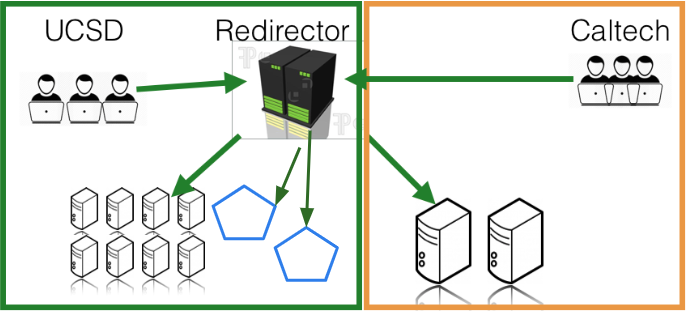}
\caption{Current Setup for SoCal deployment. The blue pentagons are XCache services deployed using kubernetes}
\label{currentSetup}       % Give a unique label
\end{figure}

\subsection{Monitoring}

Currently there are three sources of monitoring data surrounding this service:
\begin{enumerate}
    \item The job monitoring data from the Global pool HTCondor classAds that can be accessed through MONIT \cite{monit}
    \item The XRootD monitoring data form the XCache servers accessible through monit at CERN and GRACC \cite{GRACC}
    \item The hardware probes installed in the servers
\end{enumerate}

The first one allows us to have a view of the Cache from the jobs perspective and compute metrics like failure rate, average read speed and CPU efficiency.  The second allows us to gather information about the data accesses from the cache perspective and calculate metrics like total data delivered, working set size, unique reads, etc. Finally, the third source gives us more standard metrics of the host like the load of the system or the network traffic.

Making aggregations of this monitoring data over large periods of time is of great interest for the operators of this kind of service and could be incredibly time consuming and sometimes impossible given the data retention policy of the current monitoring systems. Based on these observations, Monicron was born as part of the TUDA project \cite{tuda}. Monicron is a new monitoring system designed to calculate and store aggregations of complex metrics, from both the XRootD and the Global pool jobs monitoring data, over defined intervals of time. The ultimate goal of Monicron is to provide a simplified view of the health and performance of the cache. The development of this system is complete and it is anticipated to be fully operational at the end of February, 2020.

%TODO fill out for Diego and Jonathan on the monitoring

\section{Cache Performance}
\begin{figure}[h]
% to insert the figure file..
\centering
\sidecaption
\includegraphics[width=8cm,clip]{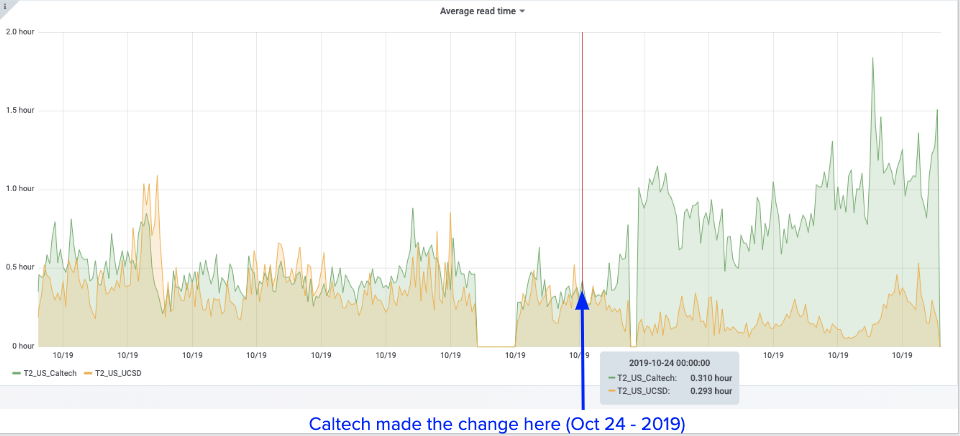}
\caption{Average Read Time for analysis jobs using MINIAOD at Caltech and UCSD}
\label{AvgReadTime}       % Give a unique label
\end{figure}

In order for us to get a better measurement of the benefits of installing the Xcache infrastructure we performed an experiment. During a month we changed the configuration of Caltech's jobs to access the MINIAOD data tier not using the cache but using local Hadoop (if files were available) or AAA. The results can be seen in fig \ref{AvgReadTime}, we find out that the average read time for jobs not using the cache was five times higher than using the cache (the "control group" UCSD and Caltech received the same mixed of jobs during that period of time).

\section{Conclusions and Future Work}

We have shown that by rapid allocation and deployment (via Kubernetes) of about one Petabyte of disk that can receive about seventy percent of CMS analyses jobs while increasing average read time speeds up to five times. The future directions on the caching deployments will be three legged. Researching different algorithms for deciding what to write and to delete to the local cache hopefully profiting of the information richness of CMS Logical file names, second on the consistency checks at the cache level (what is stored in the cache can be corrupted, over time, disk errors or even network errors) and in depth studying the overall impact of the the deployment model of containers versus bare metal in the performance of the caches and time and money savings of a DevOps model in full operation.

\section*{Acknowledgement}
The authors would like to thank the different funding agencies for this work, in particular the National Science Foundation through the following grants: OAC-1541349, MPS-1148698, OAC-1836650, MPS-1624356, OAC-1826967.
%
% BibTeX or Biber users please use (the style is already called in the class, ensure that the "woc.bst" style is in your local directory)
% \bibliography{name or your bibliography database}
%
\bibliography{Xcache.bib}

\begin{thebibliography}{9}

\bibitem{cmsDeployAAA}
K.B. for~the Cms~Collaboration, Journal of Physics: Conference Series
  \textbf{513}, 042005 (2014)

\bibitem{xrootdcms}
L.~Bauerdick, D.~Benjamin, K.~Bloom, B.~Bockelman, D.~Bradley, S.~Dasu,
  M.~Ernst, R.~Gardner, A.~Hanushevsky, H.~Ito et~al., Journal of Physics:
  Conference Series \textbf{396}, 042009 (2012)

\bibitem{dorigo2005xrootd}
A.~Dorigo, P.~Elmer, F.~Furano, A.~Hanushevsky, WSEAS Transactions on Computers
  \textbf{1} (2005)

\bibitem{xcache}
L.~Bauerdick, K.~Bloom, B.~Bockelman, D.~Bradley, S.~Dasu, J.~Dost,
  I.~Sfiligoi, A.~Tadel, M.~Tadel, F.~Wuerthwein et~al., \emph{XRootd,
  disk-based, caching proxy for optimization of data access, data placement and
  data replication} (2014), Vol. 513

\bibitem{FederatedCache}
E.~Fajardo, A.~Tadel, M.~Tadel, B.~Steer, T.~Martin, F.~Würthwein, Journal of
  Physics: Conference Series \textbf{1085}, 032025 (2018)

\bibitem{globalpool}
J.~Balcas, S.~Belforte, B.~Bockelman, D.~Colling, O.~Gutsche, D.~Hufnagel,
  F.~Khan, K.~Larson, J.~Letts, M.~Mascheroni et~al., Journal of Physics:
  Conference Series \textbf{664}, 062031 (2015)

\bibitem{monit}
A.~Aimar, A.~Corman, P.~Andrade, J.~Fernandez, B.~Bear, E.~Karavakis,
  D.~Kulikowski, L.~Magnoni, EPJ Web of Conferences \textbf{214}, 08031 (2019)

\bibitem{GRACC}
K.~Retzke, D.~Weitzel, S.~Bhat, T.~Levshina, B.~Bockelman, B.~Jayatilaka,
  C.~Sehgal, R.~Quick, F.~Wuerthwein, Journal of Physics: Conference Series
  \textbf{898}, 092044 (2017)

\bibitem{tuda}
J.~Guiang, \emph{jkguiang/tuda: Zenodo release} (2020),
  \urlstyle{tt}\url{https://doi.org/10.5281/zenodo.3636224}

\end{thebibliography}

\end{document}